# Electric field tunable multi-state tunnel magnetoresistances in 2D van der Waals magnetic heterojunctions

B. Liu[1], X. X. Ren[1], Xian Zhang[2,a)], Ping Li[1], Y. Dong[3,a)], and Zhi-Xin Guo[1,a)]

[1]State Key Laboratory for Mechanical Behavior of Materials, Center for Spintronics and Quantum System, School of Materials Science and Engineering, Xi'an Jiaotong University, Xi'an, Shaanxi, 710049, China.

[2]Shaanxi Key Laboratory of Surface Engineering and Remanufacturing College of Mechanical and Materials Engineering Xi'an University Xi'an 710065, China.

[3]State Key Laboratory of Military Stomatology & National Clinical Research Center for Oral Diseases & Shaanxi Key Laboratory of Stomatology, Department of Prosthodontics, School of Stomatology, The Fourth Military Medical University, Xi'an 710032, China

Authors to whom correspondence should be addressed: zhangxian1@outlook.com; dongyanfmmu@qq.com; zxguo08@xjtu.edu.cn

**Abstract**

Magnetic tunnel junction (MTJ) based on van der Waals (vdW) magnetic layers has been found to present excellent tunneling magnetoresistance (TMR) property, which has great potential applications in field sensing, non-volatile magnetic random access memories, and spin logics. Although MTJs composed of multilayer vdW magnetic homojunction have been extensively investigated, the ones composed of vdW magnetic heterojunction is still to be explored. Here we use first-principles approaches to reveal that the magnetic heterojunction MTJs have much more distinguishable TMR values than the homojunction ones. In the MTJ composed of bilayer $CrI_3$/bilayer $Cr_2Ge_2Te_6$ heterojunction, we find there are eight stable magnetic states, leading to six distinguishable electronic resistances. As a result, five sizable TMRs larger than 300% can be obtained (the maximum TMR is up to 620,000%). Six distinguishable memories are obtained which is two times larger than that of a four-layered homojunction MTJ. The underlying relationships among magnetic state, spin-polarized band structures, and transmission spectrums are further revealed to explain the multiple TMR values. We also find that the magnetic states and thus TMRs can be efficiently modulated by an external electric field. This study opens an avenue to the design of high-performance MTJ devices based on vdW heterojunctions.

During the past two decades, the magnetoresistance (MR) effect and its applications have been extensively studied, which gives rise to the rapidly developing field of spintronics [1-4]. Particularly, the tunneling magnetoresistance (TMR) effect provides means for all-electrical readout in magnetic random access memory (MRAM) devices [5-8]. As an application of TMR, magnetic tunnel junctions (MTJs) have become the leading devices for field sensing, non-volatile MRAM and spin logic applications [7, 8, 11-13].

MTJ is a component consisting of a ferromagnet/spacer/ferromagnet stack, usually with a TMR value of several hundred or even thousand [7-10]. Traditionally, although four distinct stable magnetic states are possible in an MTJ such as CoFeB/MgO/CoFeB, (↑↑,↓↑,↓↑,↓↓) (arrows indicate the magnetization directions of the two magnetic layers), magnetoresistive sensing can only distinguish between two resistance states, i.e., parallel and anti-parallel [14]. Recently, the discovery and application of two-dimensional (2D) van der Waals (vdW) materials in MTJ (denoted as vdW MTJ) has attracted great attention [15-25]. An advance of such MTJ is that the spacer which is usually a thin insulator is not necessarily required, because the weak vdW interlayer interaction can induce a tunneling barrier. Notably, the multi-state tunnel magnetoresistances have been realized by using multilayer vdW ferromagnets in MTJ [15,16]. For example, three resistance states with giant TMR had been observed in an MTJ consisting of a four-layer $CrI_3$ homojunction [16]. Moreover, wang et al. realized two/three resistance states in a homojunction of two/three $Fe_3GeTe_2$ nanoflakes [20].

Despite the extensive studies on MTJs composed of vdW magnetic homojunction (denoted as homojunction MTJs), the ones composed of magnetic heterojunctions (denoted as heterojunction MTJs) have been rarely explored. Since the heterojunction geometry generally leads to additional symmetry breaking, much more distinguishable magnetoresistances are expected in a heterojunction MTJ than that in a homojunction one. Thus, unveiling the TMR properties of heterojunction MTJs is of great importance to the development of high-performance spintronic devices.

By using the first-principles calculations, here we reveal distinguishable multi-state TMRs of vdW magnetic heterojunction composed of bilayer $CrI_3$ and bilayer $Cr_2Ge_2Te_6$. We find that the heterojunction can hold eight distinguishable magnetic states, which is more than two times larger than that of a four-layered magnetic homojunction. As a result, five sizable TMRs larger than 300% can be obtained, with the maximum TMR up to 620,000%. We also find that magnetic states of both bilayer $CrI_3$ and bilayer $Cr_2Ge_2Te_6$ can be efficiently tuned by an external electric field, showing that the magnetoresistances can be easily controlled via electric fields.

The calculation method is shown in Supplemental Material. The calculated in-plane lattice constants of $CrI_3$ and $Cr_2Ge_2Te_6$ are 7.00 Å and 6.92 Å, respectively, which are in excellent agreement with previous studies [26-28]. As shown in Figs. 1(b) and 1(d), in AFM (FM) state both bilayer $Cr_2Ge_2Te_6$ and bilayer $CrI_3$ are semiconductors with band gaps of 0.61 (0.40) eV and 1.14 (1.04) eV, respectively. Moreover, both of their conduction band minimum (CBM) is contributed by a pure spin-polarized state, indicating that they are ideal candidates for MTJ devices.

The combination of bilayer $Cr_2Ge_2Te_6$ and bilayer $CrI_3$ forms a four-layered magnetic heterojunction. Here we considered several typical stacking configurations (see Fig. S1), and found the configuration with para-position of I and Ge atoms at the interface is the most stable [Fig. S1(b)]. The $Cr_2Ge_2Te_6$-$CrI_3$ interlayer distance is 3.37 Å with cohesive energy of about 17 meV/Å$^2$, showing the nature of vdW interface coupling. We have also calculated the magnetocrystalline anisotropy energy (MAE) of heterojunction, which is defined as the energy difference with magnetization direction between [001] and [010] directions. The MAE of the heterojunction is -3.93 meV, showing the nature of perpendicular magnetization in the heterojunction. This result implies that the intrinsic spin-polarized electronic properties of $CrI_3$ and $Cr_2Ge_2Te_6$ can be preserved in the heterojunction.

According to the symmetry analysis, the four-layered magnetic heterojunction has eight distinguishable magnetic states, that is, ↑↑↑↑ (FM), ↑↓↑↓ (AFM-I), ↑↓↓↑(AFM-II), ↑↑↓↓ (AFM-III), ↑↑↑↓ (FIM-I), ↑↑↓↑ (FIM-II), ↓↑↑↑ (FIM-III), ↑↓↑↑ (FIM-IV) [Fig. 1(e)]. The calculated total energies of the eight states are shown in Table S1, where the maximum energy difference is within 20 meV. Figure 2 additionally shows the projected energy bands of the eight possible states. A common characteristic is that the band-structure morphologies of bilayer $CrI_3$ and bilayer $Cr_2Ge_2Te_6$ change little after stacking into a heterojunction, owing to the weak vdW interlayer coupling between $Cr_2Ge_2Te_6$ and $CrI_3$. As a result, the spin-projected band structures of the AFM-I state ↑↓↑↓ and AFM-II state ↑↓↓↑ are analogous. Similarly, it is not surprising there is little band-structure difference between FIM-I ↑↑↑↓ and FIM-II ↑↑↓↑ states, and that between FIM-III ↓↑↑↑ and FIM-IV ↑↓↑↑ states. It is worth noting that similar band structures do not guarantee a similar spin-polarized transport property. As shown in the calculated TMR results below, despite there being only five distinguishable band structures, five distinguishable TMR values can still be produced by the eight magnetic states.

The schematic diagram of the MTJ device is shown in Fig. 1(f), where the $Cr_2Ge_2Te_6$/$CrI_3$ heterojunction is sandwiched between two multilayer graphene electrodes. The ATK software was used to calculate the spin-polarized transport property. Fig. 3 shows the calculated spin-resolved transmission spectra of the eight magnetic states in an energy range of [-2, 2] eV. A common feature is that the transmission spectra difference between spin-up and spin-down electrons of all the eight magnetic states is very small below the Fermi level ($E_F$), whereas it becomes significantly large above $E_F$. This phenomenon can be understood from the spin-resolved band structures. As shown in Fig. 2, the energy bands of both spin-up and spin-down electrons in the heterojunction largely overlap below $E_F$, so there is almost no isolated spin-polarized state in an energy range of [-2, 0] eV. However, above $E_F$, the isolated spin-polarized states of $CrI_3$ appear, which can lead to a significant difference in the spin-resolved transmission spectra. Note that the transmission spectra profiles of the eight magnetic states are distinguishable from each other above $E_F$, owing to the multiplex spin-polarized band alignments in the $Cr_2Ge_2Te_6$/$CrI_3$ heterostructure. Furthermore, the transmission spectra difference has a strong dependence on the interlayer magnetic order of bilayer $CrI_3$. When the bilayer $CrI_3$ is in FM interlayer coupling, there is a significant transmission difference between spin-up and spin-down electrons. Whereas, when the $CrI_3$ is in the AFM interlayer coupling state, the transmission difference is tiny. This phenomenon mainly resulted from the more compatible energy levels of FM $CrI_3$ with that of $Cr_2Ge_2Te_6$, which leads to a large energy band overlap region in the conduction bands (Fig. 2).

This plentiful transmission spectra means that multiple TMR values can be obtained in the MTJ device, which motivated us to explore the I-V curve characteristics of all eight magnetic states. At a bias voltage $V_b$, the spin-resolved current can be calculated as follows

$$I_\sigma(V_b) = \frac{e}{h}\int_{-\infty}^{+\infty} dE[f(E,\mu_L) - f(E,\mu_R)]T_\sigma(E,V_b) \qquad (1)$$

where $f(E,\mu_L)$ and $f(E,\mu_R)$ are the Fermi-Dirac distribution of the left and right electrodes, respectively. $\mu_L$ and $\mu_R$ are the electrochemical potential of the left and right electrodes, respectively, and $T_\sigma(E)$ is the transmission probability for an electron at energy $E$ with spin $\sigma$. The linear response current originated from the equilibrium transmission spectrum at zero bias is considered in the calculations.

As shown in Fig. 4, the flour-layered magnetic heterojunction can lead to a novel distribution of I-V curves. The I-V curves can be divided into three groups. That is, the FM, FIM-III, and AFM-III states have the largest current, the FIM-I, FIM-II and AFM-I states have the smallest current, and the current of remaining FIM-IV states is in between. Note that the current difference among these magnetic states becomes more and more significant with the bias voltage increasing, with the most significant difference appearing in region of 0.8 V~1.0 V. Interestingly, different from conventional concept that the current of the FM state is the largest [1-5], the FIM-III and AFM-III states have the largest current in the low (<0.6 V) and high (> 0.7 V) bias-voltage regions, respectively. The variation of I-V curves of these magnetic states can be understood from the transmission spectra shown in Fig. 3, i.e., the state with a larger current has a larger T(E) in energy range of [-$V_b$/2, $V_b$/2]. Particularly, a direct comparison of T(E) for AFM-III, FIM-III, and FM states which have very similar current values is shown in Fig. S2. One can see that the larger current value under $V_b$ corresponds to a larger integration of T(E) in [-$V_b$/2, $V_b$/2].

To reveal the intrinsic correlations between the T(E) and magnetic states, we further calculated the projected local density of states (PLDOS) at zero bias along the MTJ device. As marked by the black lines in Fig. 5, the shape of the tunneling barrier in the magnetic heterojunction region (central region) strongly depends on the interlayer magnetic configuration. In the following, we will show that the shape of the tunneling barrier directly correlates to the electric resistance, and thus the I-V curves. According to the Wentzel-Kramers-Brillouin (WKB) formula, the tunneling transmission function T can be written as [19, 30]

$$T = \exp\left\{-4\int \frac{dx\sqrt{2\mu[U - E_F + B(x)\sigma]}}{\hbar}\right\} \quad (2)$$

where $U - E_F$ represents the tunneling barrier height in the magnetic system, $\mu$ is the effective mass of electrons, $B(x)$ is the electric field related to $x$, and $\hbar$ is Planck's constant. Note that $U - E_F + B(x)\sigma$ is equivalent to the height of the tunneling barrier with respect to $E_F$, which is defined as h hereafter. The tunneling barrier in the central region can be divided into n discrete rectangles, so that

$$T \approx \exp\left(-\frac{4\sqrt{2\mu}}{\hbar}\sum_{i=1}^{n}\sqrt{d}\sqrt{dh_i}\right) = \exp\left(-\frac{4\sqrt{2\mu}}{\hbar}\sqrt{d}\sum_{i=1}^{n}\sqrt{s_i}\right) \quad (3)$$

where $d$ is the center-to-center distance of two adjacent magnetic layers, $i$ represents the index of magnetic layers, and $s_i=dh_i$ with $h_i$ being the height of ith layer's tunneling barrier. From the above transmission formula, the difference in T mainly lies in S=$\sum s_i$ in this study, that is, the area of potential barrier, which directly corresponds to the transmission barrier shapes in the center region.

According to the spin-polarized PLDOS shown in Fig. 5, if one simply considers a spin-polarized electron transporting across the four magnetic layers, e.g., in an order of $Cr_2Ge_2Te_6$-$Cr_2Ge_2Te_6$-$CrI_3$-$CrI_3$, three features on the transmission barrier can be obtained: (1) The transmission barrier is low/high when an electron transporting across $CrI_3$ layers with the magnetization direction parallel/antiparallel to its spin-polarization direction, due to the appearance/absence of isolated spin-polarized electronic states nearby $E_F$ as indicated by the band structures in Fig. 2; (2) The transmission barrier is much lower when an electron transporting across $Cr_2Ge_2Te_6$ layer than that across $CrI_3$ layer with opposite spin polarization, owing to the much smaller spin dependent band gaps in $Cr_2Ge_2Te_6$; (3) The transmission barriers between two $Cr_2Ge_2Te_6$ layers are basically identical (except for the FIM-IV and AFM-I states with small transmission barrier difference between two $Cr_2Ge_2Te_6$ layers) in both FM and AFM states. This is because FM and AFM states of bilayer $Cr_2Ge_2Te_6$ have similar band gaps with either spin-up or spin-down electrons (Fig. 2), which is responsible for electron transmission. Note that such band gaps can be affected by the interlayer coupling from $CrI_3$ to a certain extent, which depends on the specific magnetic states. As a result, the transmission barrier across the bilayer $Cr_2Ge_2Te_6$ slightly depends on the magnetic states of the heterojunction.

The variation of the transmission barrier can lead to a significant change in the area of potential barrier S, which is inversely proportional to transmission probability. Note that the total current is contributed by both spin-up and spin-down electrons, whereas only the one with a larger transmission probability (smaller electronic resistance) is expected to contribute the most from the viewpoint of the two-current model [29, 31, 32]. To this end, via the variation of S for a spin polarization with a larger transmission probability (Fig. 5), one can naturally understand the relationship between total current and PLDOS for all eight magnetic states. As shown in Fig. 5, four magnetic states (FM, AFM-III, FIM-III, FIM-IV) have obviously smaller S than the remaining four (AFM-I, AFM-II, FIM-I, FIM-II), due to the absence of high transmission barrier induced by $CrI_3$ layers. This result means that the former four states have smaller electronic resistances and thus larger currents, which is in good agreement with the calculated I-V curves shown in Fig. 4. It is noticed that the AFM-III state has the smallest S (spin-down electrons) among all the magnetic states, and its PLDOS above the potential energy barrier (0.35 eV) is also the largest. This feature explains why it has the largest current value with $V_b$ >0.7 V. Since S is a product of PLDOS, the above results confirm that the electronic resistance of a multilayer magnetic vdW heterojunction strongly correlates to its microscopic band structures, which are determined by the magnetic state. In addition, the PLDOS of all the eight magnetic states with the spin polarization of smaller transmission probability is shown in Fig. S3, most of which have obviously larger S than that in Fig. 5.

To further unveil the applicability of magnetic vdW heterojunction in the MTJ device, we calculated the bias-voltage dependent TMRs with different magnetic states, where AFM-I was adopted as the reference state. The TMR is defined as

$$TMR = \left| \frac{I_x - I_{\uparrow\downarrow\uparrow\downarrow}}{I_{\uparrow\downarrow\uparrow\downarrow}} \right| \times 100\% \qquad (4)$$

where $I_x$ represents an electronic current of the magnetic states except for AFM-I. As shown in Fig. 6, eight magnetic states give rise to six distinguishable TMR curves, where two features can be obtained: (1) almost all TMR values (except for the FIM-I and AFM-II states) increase first and then decrease with the bias voltage increasing, which reaches the maximum at about 0.8 V; (2) The TMR values become the most distinguishable at 0.8 V. It is noted that the TMR values of all the magnetic states (except for the FIM-I and AFM-II states) are quite sizable at 0.8 V, i.e., 300%, 17000%, 35000%, 110000%, and 620000% for FM-II, FIM-IV, FM, FIM-III, and AFM-III states, respectively. This result clearly shows that six distinguishable memories can be realized via a 4-layered magnetic heterojunction, which is two times larger than the homojunction case [16].

Finally, we discuss the electric field controllability on the magnetic state of the vdW heterojunction. We propose that the magnetic state can be efficiently modulated by using the one-dimensional (1D) contact technique. As shown in the attached Fig. S4 (a), the very thin monolayer/multilayer graphene can be used to contact the edge of magnetic layers. In this way, various electric fields can be applied to $CrI_3$ bilayers, $Cr_2Ge_2Te_6$ bilayers, and $CrI_3$-$Cr_2Ge_2Te_6$ bilayers by changing the external voltage $V_1$, $V_2$, and $V_3$, respectively. In addition, we have calculated the electric field dependent energy difference between AFM and FM states for the above three bilayer structures. As shown in Figs. S4(b)-S4(d), efficient magnetic state switching can be realized under the electric field effect for all three bilayer structures. Note that in the bilayer $Cr_2Ge_2Te_6$, the switching is realized with an interlayer distance of 2.90 Å, which is a little smaller than its equilibrium distance (3.26 Å) [33]. This result means that efficient switching among the eight magnetic states can be realized by the 1D contact technique [34].

In conclusion, based on the MTJ composed of bilayer $CrI_3$/bilayer $Cr_2Ge_2Te_6$ heterojunction, we have theoretically demonstrated that the multilayer vdW magnetic heterojunctions can make much more plentiful TMR values than the widely studied vdW homojunctions. The DFT calculations show there are eight stable magnetic states in the four-layered magnetic heterojunction, more than two times larger than the homojunction case. Spin-polarized transport calculations further show that the eight magnetic states can lead to six distinguishable electronic resistances. Despite the weak interlayer interactions, the interlayer magnetic configurations can effectively influence the spin-polarized band alignments and band gaps near $E_F$. The variation of band structures can lead to significant change in transmission barrier and thus the transmission probability, which is responsible to the formation of multi-state magnetic resistances. Consequently, five distinguishable TMRs that are larger than 300% are obtained, with the maximum value up to 620,000%. Moreover, the magnetic states and thus TMRs can be efficiently modulated by an external electric field. This result shows that the multi-state TMRs in such heterojunction MTJ device can be effectively realized with low operation energy, which has great potential applications in the electrically coupled spintronic devices.

**Supplementary Material**

See the supplementary material for the calculation method and additional results.

We acknowledge financial support from the Ministry of Science and Technology of the People´s Republic of China (Grant No. 2022YFA1402901) and the Natural Science Foundation of China (Grant Nos. 12074301, 12004295). We gratefully acknowledge the computational resources provided by the HPCC platform of Xi'an Jiaotong University.

**References:**
1. M. N. Baibich, J. M. Broto, A. Fert, F. Nguyen Van Dau, F. Petroff, P. Etienne, G. Creuzet, A. Friederich, and J. Chazelas, Phys. Rev. Lett. 61, 2472 (1988).
2. G. Binasch, P. Grünberg, F. Saurenbach, and W. Zinn, Phys. Rev. B 39, 4828(R) (1989).
3. S. Yuasa, T. Nagahama, A. Fukushima, Y. Suzuki, and K. Ando, Nat. Mater. 3, 868 (2004).
4. I. Žutić, J. Fabian, and S. Das Sarma, Rev. Mod. Phys. 76, 323 (2004).

5. S. Ikeda, J. Hayakawa, Y. Ashizawa, Y. M. Lee, K. Miura, H. Hasegawa, M. Tsunoda, F. Matsukura, and H. Ohno, Appl. Phys. Lett. 93, 082508 (2008).
6. F. Schleicher, U. Halisdemir, D. Lacour, M. Gallart, S. Boukari, G. Schmerber, V. Davesne, P. Panissod, D. Halley, H. Majjad, Y. Henry, B. Leconte, A. Boulard, D. Spor, N. Beyer, C. Kieber, E. Sternitzky, O. Cregut, M. Ziegler, F. Montaigne, E. Beaurepaire, P. Gilliot, M. Hehn and M. Bowen, Nat. Commun. 5, 4547 (2014).
7. N. Maciel, E. Marques, L. Naviner, Y. Zhou, and H. Cai, Sensors 20, 121 (2019).
8. B. Jinnai, K. Watanabe, S. Fukami, and H. Ohno, Appl. Phys. Lett. 116, 160501 (2020).
9. D. Waldron, V. Timoshevskii, Y. Hu, K. Xia, and H. Guo. Phys. Rev. Lett. 97, 226802 (2006).
10. T. Scheike, Q. Xiang, Z. Wen, H. Sukegawa, T. Ohkubo, K. Hono, and S. Mitani, Appl. Phys. Lett. 118, 042411 (2021).
11. M. Wang, W. Cai, D. Zhu, Z. Wang, J. Kan, Z. Zhao, K. Cao, Z. Wang, Y. Zhang, T. Zhang, C. Park, J.-P. Wang, A. Fert, and W. Zhao, Nat. Electron. 1, 582 (2018).
12. S. Shi, Y. Ou, S. V. Aradhya, D. C. Ralph, and R. A. Buhrman, Phys. Rev. Appl. 9, 011002 (2018).
13. A. Manchon, J. Železný, I. M. Miron, T. Jungwirth, J. Sinova, A. Thiaville, K. Garello, and P. Gambardella, Rev. Mod. Phys. 91, 035004 (2019).
14. C. O. Avci, M. Mann, A. J. Tan, P. Gambardella, and G. S. D. Beach, Appl. Phys. Lett. 110, 203506 (2017).
15. T. Song, Q. Sun, E. Anderson, C. Wang, J. Qian, T. Taniguchi, K. Watanabe, M. A. McGuire, R. Stöhr, D. Xiao, T. Cao, J. Wrachtrup, and X. Xu, Science 360, 1214 (2018).
16. T. Song, M. W. Tu, C. Carnahan, X. Cai, T. Taniguchi, K. Watanabe, M. A. McGuire, D. H. Cobden, D. Xiao, W. Yao, and X. Xu, Nano Lett. 19, 915 (2019).
17. L. Zhang, T. Li, J. Li, Y. Jiang, J. Yuan, and H. Li, J. Phys. Chem. C 124, 27429 (2020).
18. Y. Su, X. Li, M. Zhu, J. Zhang, L. You, and E. Y. Tsymbal, Nano Lett. 21, 175 (2021).
19. J. Yang, S. Fang, Y. Peng, S. Liu, B. Wu, R. Quhe, S. Ding, C. Yang, J. Ma, B. Shi, L. Xu, X. Sun, G. Tian, C. Wang, J. Shi, J. Lu, and J. Yang, Phys. Rev. Appl. 16, 024011 (2021).
20. C. Hu, D. Zhang, F. Yan, Y. Li, Q. Lv, W. Zhu, Z. Wei, K. Chang, and K. Wang, Sci. Bull. 65, 1072 (2020).
21. W. Zhu, H. Lin, F. Yan, C. Hu, Z. Wang, L. Zhao, Y. Deng, Z. R. Kudrynskyi, T. Zhou, Z. D. Kovalyuk, Y. Zheng, A. Patanè, I. Žutić, S. Li, H. Zheng, and K. Wang, Adv. Mater. 33, 2104658 (2021).
22. D. Li, T. Frauenheim, and J. He, ACS Appl. Mater. Interfaces 13, 36098 (2021).
23. H. Zhou, Y. Zhang, and W. Zhao, ACS Appl. Mater. Interfaces 13, 1214 (2021)
24. L. Cao, X. Deng, G. Zhou, S. Liang, C. V. Nguyen, L. K. Ang, and Y. S. Ang, Phys. Rev. B 105, 165302 (2022).
25. B. Wu, J. Yang, R. Quhe, S. Liu, C. Yang, Q. Li, J. Ma, Y. Peng, S. Fang, J. Shi, J. Yang, J. Lu, and H. Du, Phys. Rev. Appl. 17, 034030 (2022).
26. R. Xu and X. Zou, J. Phys. Chem. Lett. 11, 3152 (2020).
27. Y. F. Li, W. Wang, W. Guo, C. Y. Gu, H. Y. Sun, L. He, J. Zhou, Z. B. Gu, Y. F. Nie, and X. Q. Pan, Phys. Rev. B 98, 125127 (2018).
28. P. Li, X. Zhou and Z. Guo, npj Comput. Mater. 8,20 (2022).
29. M. Julliere, Phys. Lett. A 54, 225 (1975).

30. Y. B. Band and Y. Avishai, Quantum Mechanics with Applications to Nanotechnology and Information Science (Academic Press, Amsterdam, 2013), pp. 303–366.
31. N. F. Mott, Adv. Phys. 13, 325 (1964).
32. S. Shen, P. R. Ohodnicki, S. J. Kernion, and M. E. McHenry, J. Appl. Phys. 112, 103705 (2012).
33. T. Song, Z. Fei, M. Yankowitz, Z. Lin, Q. Jiang, K. Hwangbo, Q. Zhang, B. Sun, T. Taniguchi, K. Watanabe, M. A. McGuire, D. Graf, T. Cao, J. Chu, D. H. Cobden, C. R. Dean, D. Xiao & X. Xu. Nat. Mater. 18, 1298–1302 (2019).
34. B. Huang, G. Clark, D. R. Klein, D. MacNeill, E. Navarro-Moratalla, K. L. Seyler, N. Wilson, M. A. McGuire, D. H. Cobden, D. Xiao, W. Yao, P. Jarillo-Herrero & X. Xu. Nat. Nanotechnol. 13, 544 (2018).

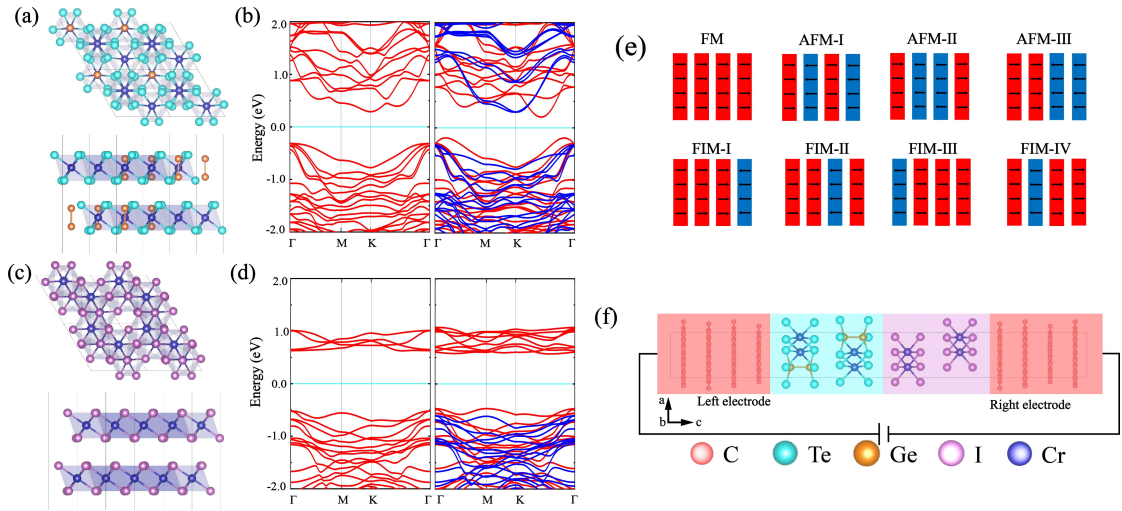

FIG. 1. Atomic structures, spin-polarized band structures of bilayer $Cr_2Ge_2Te_6$ and bilayer $CrI_3$, as well as schematic diagrams of eight stable magnetic states and MTJ device composed of the bilayer $Cr_2Ge_2Te_6$/bilayer $CrI_3$ heterojunction. (a) Top and side views of bilayer $Cr_2Ge_2Te_6$. (b) Band structures of AFM (left) and FM (right) bilayer $Cr_2Ge_2Te_6$. (c) Top and side views of bilayer $CrI_3$. (d) Band structures of AFM (left) and FM (right) bilayer $CrI_3$. (e) Schematic diagram of eight stable magnetic states, i.e., FM (↑↑↑↑), AFM-I (↑↓↑↓), AFM-II (↑↓↓↑), AFM-III (↑↑↓↓), FIM-I (↑↑↑↓), FIM-II (↑↑↓↑), FIM-III (↓↑↑↑), FIM-IV (↑↓↑↑), where the arrows in each magnetic layer indicate its magnetization direction. (f) Schematic diagram of MTJ device, with bilayer $Cr_2Ge_2Te_6$/bilayer $CrI_3$ heterojunction sandwich between two graphene electrodes. Cr, Ge, Te, I, and C atoms are depicted by the blue, yellow, cyan-blue, pink, and red balls, respectively. The red and blue lines in (b) and (d) correspond to the energy bands from spin-up and spin-down electrons, respectively.

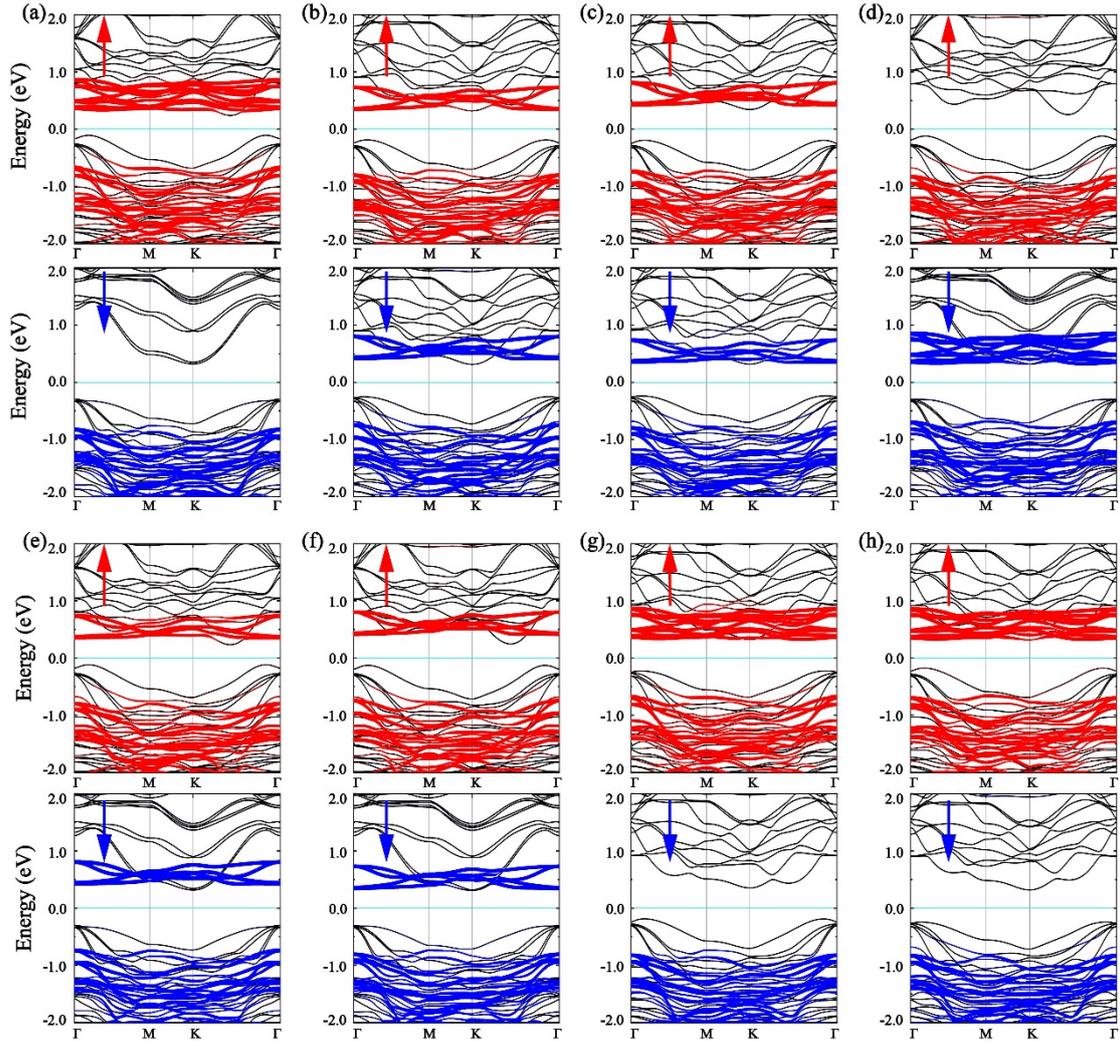

FIG. 2. Spin-polarized band structures of the eight magnetic states in bilayer $Cr_2Ge_2Te_6$/bilayer $CrI_3$ heterojunction. (a)-(h) present the band structures of FM, AFM-I, AFM-II, AFM-III, FIM-I, FIM-II, FIM-III, FIM-IV states, respectively. As indicated by the arrows, the upper and lower panels correspond to the energy bands of spin-up and spin-down electrons, respectively. In addition, the red and blue lines represent the energy bands of spin-up and spin-down electrons of $CrI_3$, respectively.

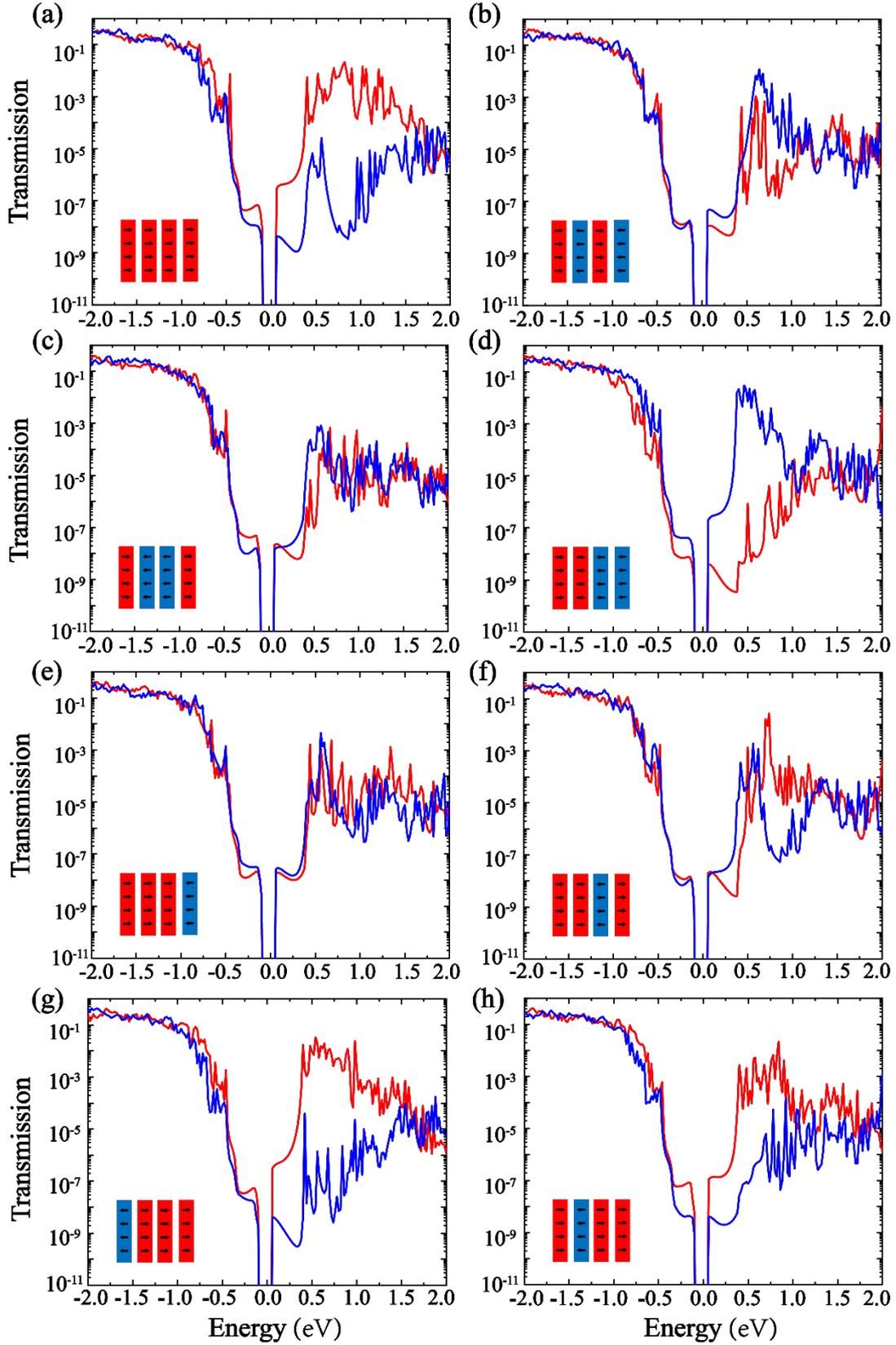

FIG. 3. Calculated spin-polarized transmission of the MTJ device with different magnetic states. (a)-(h) present the transmission spectra of FM, AFM-I, AFM-II, AFM-III, FIM-I, FIM-II, FIM-III, FIM-IV states, respectively. The red and blue lines represent the transmission of spin-up and spin-down electrons, respectively.

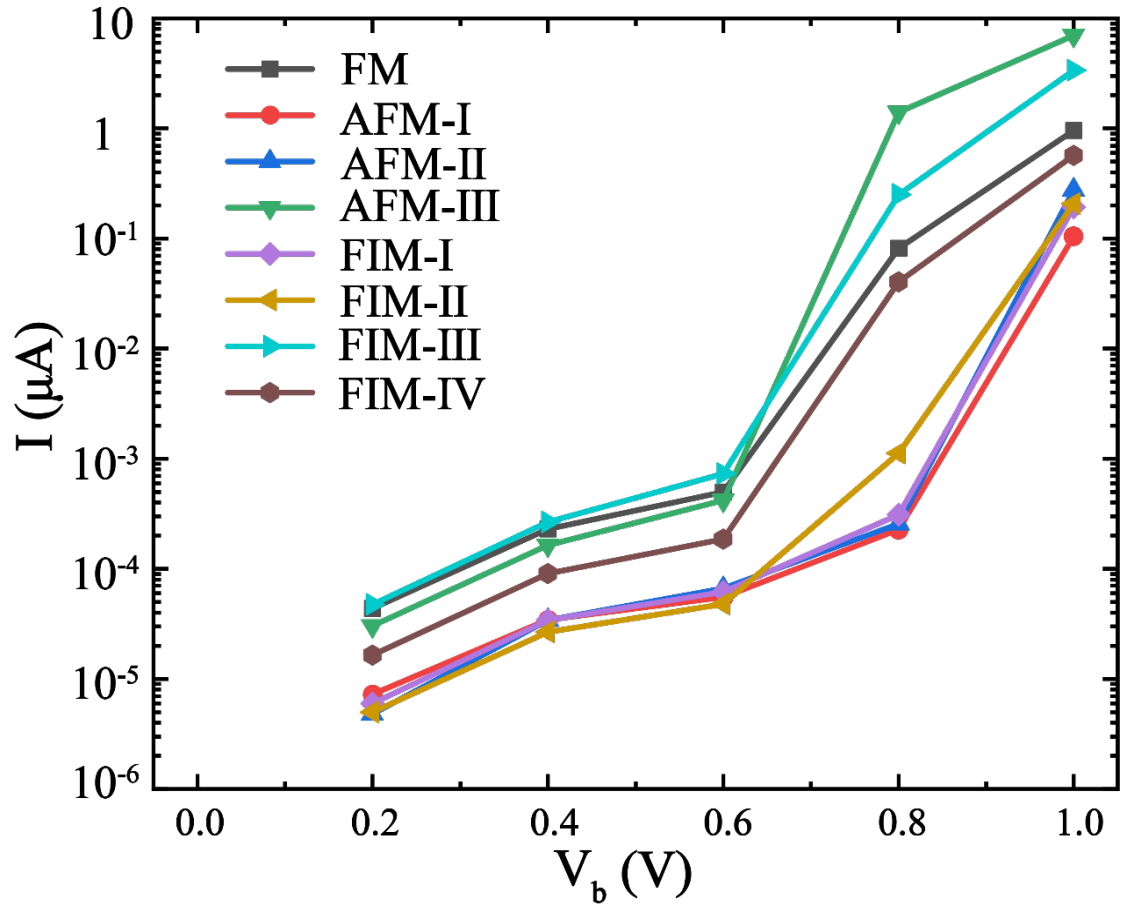

FIG. 4. Calculated I-V curves of the MTJ device with different magnetic states. The flour-layered magnetic heterojunction can lead to three groups distinguishable I-V curves.

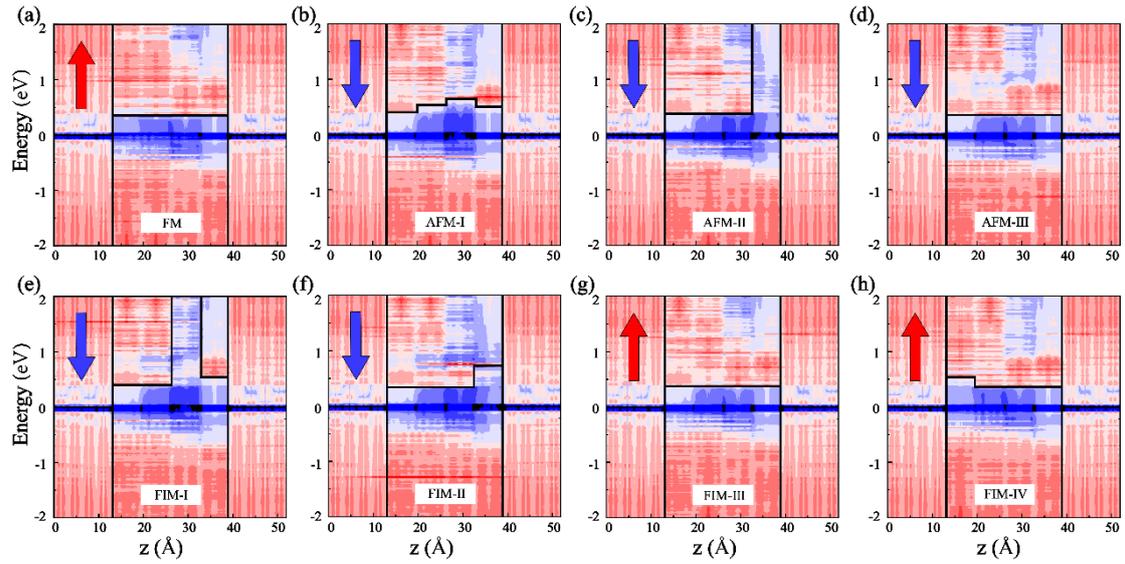

FIG. 5. Electron projected local density of states (PLDOS) of the eight magnetic states in the MTJ device. Note that each magnetic state has two PLDOS, which are contributed by spin-up and spin-down electrons, respectively. Here only the one with smaller potential barrier area is shown (spin polarization is indicated by the arrows), and the remaining one is shown in the Supplemental Material. The black lines indicate the shape of potential barrier. Note that the bilayer $Cr_2Ge_2Te_6$ ($CrI_3$) locates in the region of 11 Å - 25 Å (26 Å - 41 Å).

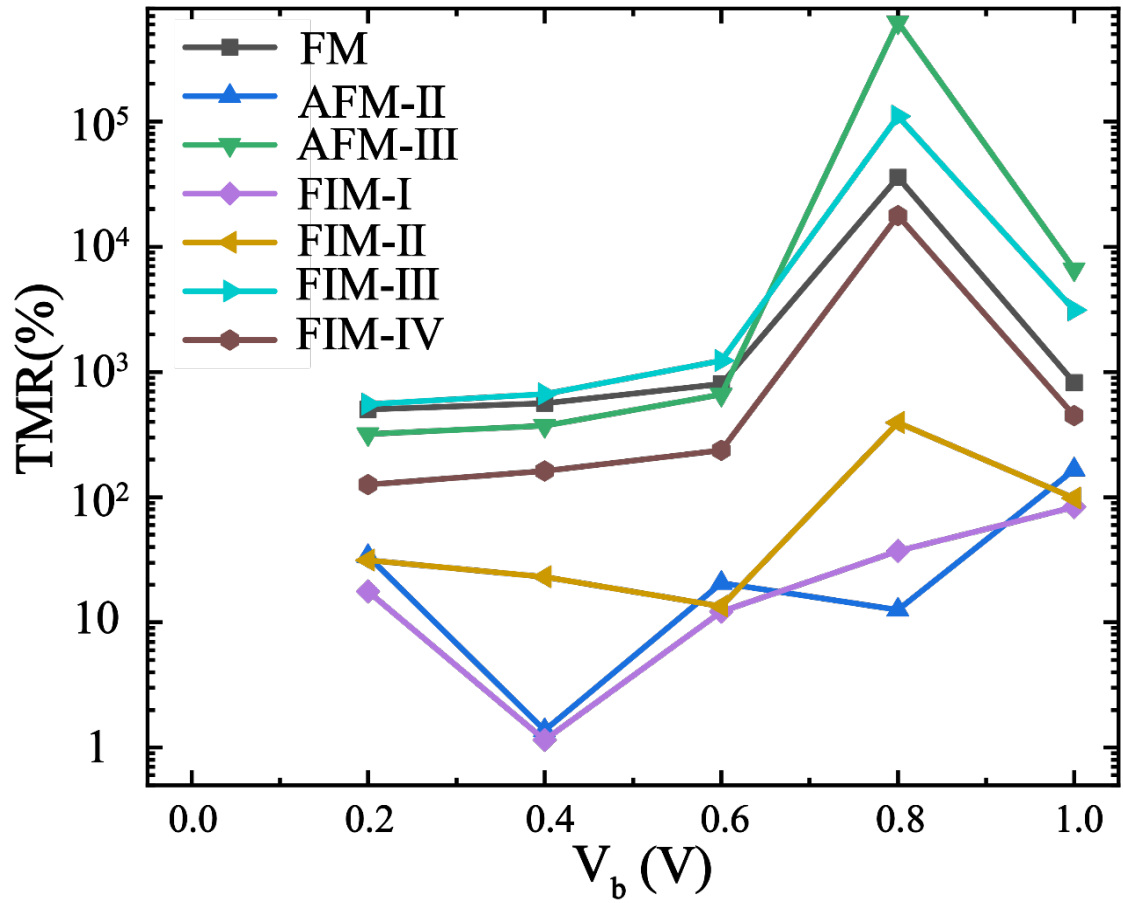

FIG. 6. Bias dependent TMRs of different magnetic states, with AFM-I state as the reference state. The TMR values become the most distinguishable at 0.8 V.